\documentclass[preprint,showpacs,preprintnumbers,amsmath,amssymb,superscriptaddress]{revtex4-1}
\usepackage{graphicx}% Include figure files
\usepackage{dcolumn}% Align table columns on decimal point
\usepackage{bm}% bold math
\usepackage{tabularx}
\usepackage{makecell}
\usepackage{braket}
\usepackage{lipsum}
\usepackage{ulem}
\usepackage{amsmath,amssymb}
\newcolumntype{M}{>{\centering\arraybackslash}m{1.85cm}}
\usepackage[export]{adjustbox}
\usepackage{longtable}
\usepackage{float}
\usepackage{xcolor}
\usepackage{color}   %May be necessary if you want to color links
\usepackage{hyperref}
\hypersetup{
	colorlinks=true, %set true if you want colored links
	linktoc=all,     %set to all if you want both sections and subsections linked
	linkcolor=blue,  %choose some color if you want links to stand out
}

\makeatletter
\newcommand{\colorcaption}[2][]{%
	\begingroup%
	\renewcommand{\@caption@fignum@sep}{ (Color online). }%
	\caption[#1]{#2}%
	\endgroup%
}
\makeatother
\newcommand{\orcid}[1]{\href{https://orcid.org/#1}{\hskip2pt\includegraphics[width=9pt]{orcid-ID.png}}}
\begin{document}

%\title{Properties of Mid-Shell Nuclei-Splittings and Isomers}
\title{ {\color{black}Spectroscopy of $^{211,213,215}$Pb isotopes and seniority properties}}
 	%\title{ Comparison of selection rules for 5 neutrons in the g$_{9/2}$ shell with that of 4 protons and 4 neutrons in the $f_{7/2}$ shell}

	\author{Larry Zamick}
    \email{Email: larry.zamick@rutgers.edu}
	\address{Department of Physics and Astronomy, Rutgers University, Piscataway, NJ. 08854 USA}
	\author{P. C. Srivastava}
     %\email{Email: praveen.srivastava@ph.iitr.ac.in}
	\address{Department of Physics, Indian Institute of Technology Roorkee, Roorkee 247667, India}

	\date{\hfill \today}
	%%%%%%%%%%%%%%%%%%%%%%%%%%%%%%%%%%%%%%%%%
	%\bibliographystyle{prsty}
	\begin{abstract}

    We consider  lead isotopes with valence neutrons in the $g_{9/2}$ shell. We especially compare $^{211}$Pb (n=3), $^{213}$Pb (n=5) and $^{215}$Pb (n=7), $^{213}$Pb being at mid-shell. Motivated by the fact the $J=21/2^+$ and J =3/2$^+$ states are pure seniority v=3 states for both n=3 and n=5, we compare the energy splitting E(J=21/2$^+$)-E(J=3/2$^+$)  for various interactions.
    %We next address the fact that the J=21/2$^+$ state is isomeric for both n=3 and n=5.
    We then discuss the position of the lowest J=3/2$^+$ state in both n=3 and n=5, although, this state has not been found experimentally. Calculations with configurations beyond g$_{9/2}$ are also considered.  We next address the fact that the J=21/2$^+$ state is isomeric for both n=3 and n=5. 
    %Finally, we draw some analogy with $^{48}$Cr  for which  the simplest configuration consist of 4 protons and 4 neutrons in the f$_{7/2}$ shell-mid-shell for both neutrons and protons.

	\end{abstract}

	\maketitle

\section{Introduction}

 Mid-shell nuclei are of interest because of their special seniority properties. As an
example let's contrast $^{211}$Pb with 3 valence $g_{9/2}$ neutrons and $^{213}$Pb with 5 valence
neutrons, the latter being at mid-shell. For n=3, all states have v=3 except for J=9/2. A
B(E2) from v=3 to v=3 is allowed but inhibited.
 For n=5 (mid-shell) we can have both v=3 and v=5 for a state of give angular
momentum-for example for J=17/2 there is one v=3 state and one v=5. These states
cannot be mixed even with a seniority non-conservation interaction. There is also a
selection rule coming into play- Electric quadrupole transitions cannot take place
between states of the same seniority. This is not true for n=3.
 On the other hand the spectrum of the pure v=3 states in $^{213}$Pb does not have to be
the same as in $^{211}$Pb. This was shown by Escuderos and  Zamick \cite{1_1}. We here focus
on the splitting of J=21/2 and J=3/2 states. Both occur only once and are pure v=3
states in both nuclei. Only with seniority conserving interactions will the spectra be the
same.
% We also draw analogies with $^{48}$Cr, a nucleus with both protos and neutrons at midshell. But we examine the isomerism of the J=21/2$^+$ state in $^{213}$Pb and see how it
%relates or does not relate to seniority.

%\section{Five neutrons in the $g_{9/2}$ shell}

%Examples of nuclei with mid-shell configuration include $^{213}$Pb (5 neutrons in the g$_{9/2}$ shell) and $^{95}$Rh (5 proton holes the g$_{9/2}$ shell ). 
%As noted by G. Racah \cite{1} and as mentioned in R.D. Lawson’s book \cite{2}, 
Racah introduced the concept of seniority \cite{1} and R.D. Lawson  stated very explicitly in his book the mid-shell selection rules for this quantum number \cite{2}. In the g$_{9/2}$ shell at mid-shell one can only have seniority mixing between states whose seniorities differ by 4 units. Examining the Bayman-Lande tables of cfp’s \cite{3} v=1
only occurs for J = 9/2$^+$. For all other J’s we only have v=3 and / or v=5 so there will be no seniority mixing for these cases.

The above topic is discussed extensively by J.J. Valiente-Dobón et al. \cite{4} in a paper titled ``Manifestation of the Berry phase in the atomic nucleus $^{213}$Pb". To quote from their abstract “The conservation of seniority is a consequence of a geometric phase associated with particle-hole conjugation which becomes observable in semi-magic nuclei where nucleons half fill the valence shell.” The geometric phase is called the Berry phase \cite{5}. 

There is now a very simple selection rule at mid-shell for electric quadrupole transitions between these seniority pure states \cite{1,2,4}. The B(E2)’s will vanish if $v_f$ = $v_i$. We can have non-zero B(E2)’s if the initial and final B(E2)’s differ by 2 units, e.g. v=3 to v=5 or v=5 to v=5 but not to from 3 to 3 or 5 to 5 with regards to B(M1), it is known that they are zero for configurations of only nucleons of one kind in a single $j$
shell.

It was noted by Zamick \cite{9} that the lowest J=3/2$^+$ states in $^{211}$Pb , $^{213}$Pb and $^{215}$Pb  have not been identified experimentally.  

%We find that with our full model space the excitation energies of these states are respectively 0.881 0.918  and 0.984 MeV.

{\color{black} For further reading, please refer following reviews, a reviews by P. M. Walker and Z. Podoly\'ak \cite{Walker},  by B.A. Brown  \cite{Brown}, by P.Van Isacker \cite{11}, by B. Maheshwari and K.Nomura \cite{12} and other books like the one by I. Talmi \cite{13}. And let’s not forget Berry \cite{5} and Bell \cite{15}.
Lawson \cite{2} relies heavily on Bell’s 1959 work on “the particle-hole conjugation operator” to
show that seniority is a good quantum number at midshell.
There are also analogous works in other regions such as the one by B. Das et al.  investigated broken seniority symmetry in the semimagic proton mid-shell nucleus $^{95}$Rh having 5 protons in the $g_{9/2}$ shell \cite{das}. Review by Racah \cite{Racah1,Racah2,Racah3} are very useful to learn theory of complex spectra.

 We have several other works that discuss isomeric states in the “lead region”.
These include the following topics, 
systematic shell-model study of Rn isotopes with A= 207 to 216 and isomeric states \cite{Bharti1}, Systematic shell model study for N= 82 and N= 126 isotones and nuclear isomers \cite{Bharti2}, systematic shell-model study of structure and isomeric states in $^{204-213}$Bi isotopes \cite{Sakshi1}, systematic shell-model study for structure and isomeric states in $^{200-210}$Po isotopes \cite{Sakshi2}, minimal theory of isomerism- Q.Q and other interactions \cite{Praveen1}, and collectivity and isomers in the Pb isotopes \cite{Praveen2,Sakshi25}.}

{\color{black} In the section II, we have reported splitting of E($21/2^+$)- E($3/2^+$). Section III reports about missing levels, the details about isomerism is reported in the section IV, and Finally we summarized our manuscript in section V.}

\begin{table*}        
\begin{center}
%\begin{threeparttable}
\caption{Energy difference between E$(21/2^+)$ - $E(3/2^+)$ using different set of interactions. In shell model calculation, with $g_{9/2}$ only means $^{208}$Pb core and neutrons in  $1g_{9/2}$ orbital only. Full means with $^{208}$Pb core and neutrons in  $1g_{9/2}$, $0i_{11/2}$, and $0j_{15/2}$ orbitals only. }
\label{tab1}
\begin{ruledtabular}
\begin{tabular}{cccc}
&$^{211}$Pb & $^{213}$Pb	& $^{215}$Pb \\[+2pt]
 \hline \\[-10pt]

State & DELTA  & & \\[+1pt]
\hline
$21/2_1^+$ &2.264 & 2.264 & 2.264 \\[+1pt]
$3/2_1^+$ & 1.759  & 1.759 & 1.759 \\[+1pt]
&Difference& & \\[+1pt]
&0.505  & 0.505 & 0.505 \\[+1pt]

  \hline \\[-10pt]
&Q.Q& & \\[+1pt]
\hline
$21/2_1^+$ & 1.606 & 1.545 & 1.606\\[+1pt]
$3/2_1^+$ &1.211  & 1.939 & 1.212 \\[+1pt]
&Difference& & \\[+1pt]
&0.394  & -0.394 & 0.394\\[+1pt]

 \hline \\[-10pt]
&$g_{9/2}$ only & & \\[+1pt]
\hline
$21/2_1^+$ & 0.709 & 0.709 & 0.709\\[+1pt]
$3/2_1^+$ &0.423  & 0.426 & 0.423 \\[+1pt]
&Difference& & \\[+1pt]
&0.286  & 0.283 & 0.286 \\[+1pt]

 \hline \\[-10pt]
&Full  & & \\[+1pt]
\hline
$21/2_1^+$ & 1.185 & 1.233 & 1.285\\[+1pt]
$3/2_1^+$ & 0.881  & 0.918 & 0.984 \\[+1pt]
&Difference& & \\[+1pt]
&0.304  & 0.315 & 0.301\\[+1pt]

\end{tabular}
\end{ruledtabular}
\vspace{-10pt}
\end{center}
%\end{threeparttable}
\end{table*}

\begin{figure}
\begin{center}
\includegraphics[width=16.50cm,height=10cm]{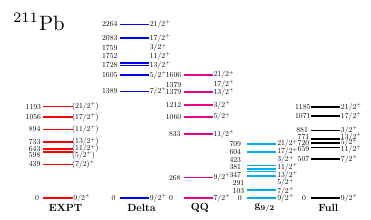}
\caption{Comparison between experimental \cite{nndc} and calculated energy levels for $^{211}$Pb isotope using different set of effective interactions.}
\label{fig1}
\end{center}
\end{figure}

\begin{figure}
\begin{center}
\includegraphics[width=16.50cm,height=10cm]{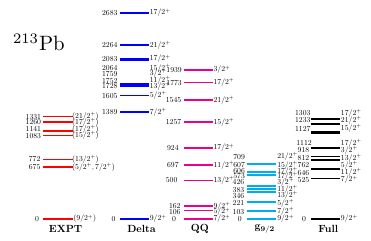}
\caption{Comparison between experimental \cite{nndc} and calculated energy levels for $^{213}$Pb isotope using different set of effective interactions.}
\label{fig2}
\end{center}
\end{figure}

\begin{figure}
\begin{center}
\includegraphics[width=16.50cm,height=10cm]{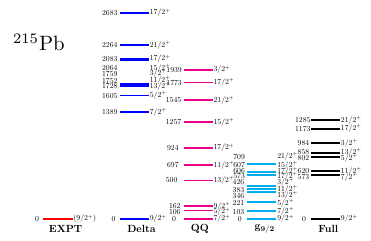}
\caption{{\color{black}Comparison between experimental \cite{nndc} and calculated energy levels for $^{215}$Pb isotope using different set of effective interactions.}}
\label{fig3}
\end{center}
\end{figure}

\section {The splitting of E($21/2^+$) – E($3/2^+$) in $^{211}$Pb, $^{213}$Pb  and $^{215}$Pb}

 In the single j shell model space $^{211}$Pb, $^{213}$Pb  and $^{215}$Pb consist of 3, 5 and 7 neutrons in the $g_{9/2}$ shell. The 7 neutrons can be regarded as 3 neutron holes as so will have identical spectra as n=3 for any interaction. For n=3  the spins J= $3/2^+$ and J = $21/2^+$ occur only once   and each has seniority v=3. They are the lowest and highest angular momenta  for the 3 particle system.  For n=5, J = 3/2$^+$ and  J = $21/2^+$ also occur only once  and have seniority v=3. So, we don’t have to worry about seniority mixing.
 As noted in ref. \cite{1_1}, in  general the value of SPLIT= E($21/2^+$)- E(3/2$^+$)  is not the same for n =3 and n=5.    The exception is a seniority conserving  interaction for which SPLIT is the same for n=3  and n=5.   The delta interaction  is a example of a seniority conserving interaction.
We show results of SPLIT in Table \ref{tab1}.
  It should be noted that the the Q.Q interaction is not seniority
conserving. Several years ago Zamick and Escuderos \cite{1_1} presented the amusing result  that  with a Q.Q interaction  the values of split  are equal and opposite for n=3 and n=5.
In Table \ref{tab1} we present the values of E($3/2^+$) , E($21/2^+$) and SPLIT for = 3, 5 and 7. They confirm what we said above about a delta interaction  (same SPLIT for n=3 and n=5) and a Q.Q interaction (equal but opposite SPLIT).
           We have also done  calculations with the KHH7B interaction \cite{10_1}. 
           To diagonalize matrices, we have used shell-model code KShell \cite{kshell}.  
          {\color{black}  The KHH7B interaction consists of total 14 orbitals out of which there are seven proton orbitals between $Z = 58-114$: $1d_{5/2}$, $1d_{3/2}$, $2s_{1/2}$, $0h_{11/2}$, $0h_{9/2}$, $1f_{7/2}$, $0i_{13/2}$ and seven neutron orbitals between $N = 100-164$: $1f_{5/2}$, $2p_{3/2}$, $2p_{1/2}$, $0i_{13/2}$, $1g_{9/2}$, $0i_{11/2}$, $0j_{15/2}$. We have completely filled the proton orbitals below Z=82 and neutron orbitals up to N=126.}
           First  we only   include  the {\color{black} neutron} $g_{9/2}$  interaction. This tells us what happens when arbitrary interaction  is used. We see that SPLIT for n=3 is the same as split n=7.  This is to be expected because in a single j shell the hole-hole interaction  is the same as the particle-particle interaction for any interaction. Note that the difference of SPLIT for n=3 vs =5 is very small.
In  the full space  there is no reason for SPLIT for n= 3 and = 7 to be the same and indeed they are not 0.304 vs 0.301. The value for n=5 is 0.315. The differences are not as dramatic as what one gets for Q.Q, possibly indication that Q.Q is not so important for particles of one kind.  The energy level diagram for $^{211}$Pb, $^{213}$Pb and  $^{215}$Pb isotopes are shown in Figs. \ref{fig1}, \ref{fig2} and \ref{fig3}, respectively.

\begin{table*}        
\begin{center}
%\begin{threeparttable}
\caption{Calculated $B(E2)$ and Half-life. In shell model calculation, with $g_{9/2}$ only means $^{208}$Pb core and neutrons in  $1g_{9/2}$ orbital only. Full means with $^{208}$Pb core and neutrons in  $1g_{9/2}$, $0i_{11/2}$, and $0j_{15/2}$ orbitals only. In the shell model calculation we have taken effective charge $e_n=0.5e$. For the half-life calculation of $^{211,213}$Pb, we have taken experimental energy differences. The experimental half-life for $21/2^+$ state in $^{211}$Pb and $^{213}$Pb are 42(7) ns and 0.26(2) $\mu s$, respectively.}
%For $g_{9/2}$ and full, we have used 
%experimental energy differences 71.2 keV and 190 keV for half-life calculation.}
\label{tab2}
\begin{ruledtabular}
\begin{tabular}{ccccc}
Interaction & Ediff$_{SM}$ (in MeV) &  Ediff$_{Expt.}$ (in MeV)	& B(E2) (in $e^2fm^4$)   &  Half-life (in sec) \\[+2pt]
 \hline %\\[-10pt]
{\bf $^{211}$Pb} & & \\[+1pt]
\hline
DELTA & & \\[+1pt]
%\hline
$21/2_1^+ \rightarrow  17/2_1^+$ & 0.182 & 0.137  &25.19  & 4.66 $\times 10^{-7}$\\[+1pt]
\hline
QQ& & \\[+1pt]
%\hline
$21/2_1^+ \rightarrow  17/2_1^+$ & 0.227 & 0.137   &25.19  & 4.66 $\times 10^{-7}$\\[+1pt]
\hline
$g_{9/2}$ only & & \\[+1pt]
%\hline
$21/2_1^+ \rightarrow  17/2_1^+$ & 0.106 & 0.137    &25.19   & 4.66  $\times 10^{-7}$\\[+1pt]
\hline
Full & & \\[+1pt]
%\hline
$21/2_1^+ \rightarrow  17/2_1^+$ & 0.113 & 0.137   &25.35   & 4.63  $\times 10^{-7}$\\[+1pt]

\hline

{\bf $^{213}$Pb} & & \\[+1pt]
\hline
DELTA & & \\[+1pt]
%\hline
$21/2_1^+ \rightarrow  17/2_1^+$ & 0.182 & 0.190 & 0.000  & \\[+1pt]
$21/2_1^+ \rightarrow  17/2_2^+$ & -0.419 & 0.0712 & 60.63  & 5.10  $\times 10^{-6}$ \\[+1pt]
%$21/2_1^+ \rightarrow $ 17/2_2^+ & -0.419 & 60.63  & 7.23  $\times 10^{-10}$\\[+1pt]
\hline
QQ& & \\[+1pt]
%\hline
$21/2_1^+ \rightarrow  17/2_1^+$ & 0.621 & 0.190  & 60.63  & 3.77  $\times 10^{-8}$\\[+1pt]
$21/2_1^+ \rightarrow  17/2_2^+$ & -0.227 & 0.0712 & 0.000 & \\[+1pt]
\hline
$g_{9/2}$ only & & \\[+1pt]
%\hline
$21/2_1^+ \rightarrow  17/2_1^+$ & 0.136 & 0.190  & 60.28  & 3.79  $\times 10^{-8}$ \\[+1pt]
%$21/2_1^+ \rightarrow $ 17/2_1^+ & 0.136 & 60.63  & 2.02  $\times 10^{-7}$\\[+1pt]
$21/2_1^+ \rightarrow  17/2_2^+$ & 0.104 & 0.0712 & 0.000 & \\[+1pt]
\hline
Full  & & \\[+1pt]
%\hline
$21/2_1^+ \rightarrow  17/2_1^+$ & 0.120 & 0.190 & 1.93 & 1.18  $\times 10^{-6}$ \\[+1pt]
$21/2_1^+ \rightarrow  17/2_2^+$ & -0.070 & 0.0712 & 47.73 & 6.48 $\times 10^{-6}$\\[+1pt]
%$21/2_1^+ \rightarrow $ 17/2_2^+ & -0.070& 47.73 & 7.06  $\times 10^{-6}$\\[+1pt]
%\hline

\end{tabular}
\end{ruledtabular}
\vspace{-10pt}
\end{center}
%\end{threeparttable}
\end{table*}

\begin{table*}        
\begin{center}
\begin{ruledtabular}
\begin{tabular}{ccccc}

{\bf $^{215}$Pb} & & \\[+1pt]
\hline
DELTA & & \\[+1pt]
%\hline
$21/2_1^+ \rightarrow  17/2_1^+$ & 0.182 & - &25.48  & 1.11 $\times 10^{-7}$\\[+1pt]
%\hline

QQ& & \\[+1pt]
%\hline
$21/2_1^+ \rightarrow  17/2_1^+$ & 0.227 & - &25.48  & 3.69 $\times 10^{-8}$\\[+1pt]
\hline
$g_{9/2}$ only & & \\[+1pt]
%\hline
$21/2_1^+ \rightarrow  17/2_1^+$ & 0.106 & -  &25.48   & 1.66  $\times 10^{-6}$\\[+1pt]

\hline
Full & & \\[+1pt]
%\hline
$21/2_1^+ \rightarrow  17/2_1^+$ & 0.111 & - &10.59   &  3.17 $\times 10^{-6}$\\[+1pt]

\end{tabular}
\end{ruledtabular}
\vspace{-10pt}
\end{center}
%\end{threeparttable}
\end{table*}

\section{ Missing levels} %{{Second last section }}
In a work “Why are those darn J=3/2$^+$ states invisible” it was noted by L.Zamick \cite{Zamick1} that
in all semi-magic nuclei that he considered there were no J=3/2$^+$ states listed in the NNDC.
This is particularly true in $^{211,213,215}$Pb.

 In the absence of experiment we go to theory. We have shown the calculated  spectra of $^{211}$Pb
and $^{213}$Pb in Figures \ref{fig1} and \ref{fig2}. There are several interactions considered but we will focus on
the full space calculation with the KHH7B interaction.
 The J=3/2$^+$ states are calculated to be at 0.881, 0.918 and 0.984 respectively (in the g$_{9/2}$
only case the excitation energies for $^{215}$Pb would be the same as for $^{211}$Pb). Note that the
J=3/2$^+$ state is at a higher energy than the lowest J=5/2$^+$ and J=7/2$^+$ states. This is an
important finding. Had the been at higher energies there could be a cascade down to
J=3/2$^+$ and it would not be “invisible”. It was already shown in \cite{Zamick1} that there were no
decays of neighboring nuclei into excited states of the above Pb isotopes.
Thus, all things considered we understand why the J=3/2$^+$ states are not found
experimentally.

{\color{black} The $^{208}$Pb is stable and $^{210}$Pb lives long enough (22.20 y), thus the transfer reactions can be performed on these nuclei. 
Indeed, $^{210}$Pb($t$, $d$)$^{211}$Pb reaction have been performed by
Ellegaard, Barnes and Flynn. However, they do not find any J = $3/2^+$ states \cite{Ellegaard}.}

\section{Isomerism}

\subsection{Isomerism in $^{211}$Pb}

By popular convention, an isomeric state in a nucleus is one that has a half-life longer than
1 ns. In $^{211}$Pb, the half-life of the lowest J = 21/2$^+$ state is 42 ns and in $^{213}$Pb it is 2.6 x $10^{-7}$ s
\cite{nndc}, both of these states are isomeric. To understand why this is so we have performed
calculations with various interactions. They are shown in Table \ref{tab2}.
 In this subsection we consider only $^{211}$Pb. For the first 3 interactions we work only in the $g_{9/2}$
model space. First we consider the delta interaction (which in a single j shell is the same as
the surface delta interaction SDI), next the Q.Q interaction. Then we take an interaction
which is intended for a full configuration but use it only in the $g_{9/2}$ shell ($g_{9/2}$ only). Finally
we do a calculation in a full space.
In $^{211}$Pb we have 3 neutrons in the $g_{9/2}$ space. There is only one J= 21/2$^+$ state and only one
J=17/2$^+$ state \cite{3}, both have seniority v=3. The J=21/2$^+$ state decays to the J=17/2$^+$ state
via an electric quadruple transition. Transitions between states of same seniority are not
forbidden but they are suppressed.
One striking observation of Table \ref{tab2} is that for the $g_{9/2}$ only calculations all B(E2’s) are
the same 25.19 $e^2fm^4$. This is due to as mentioned above to the uniqueness of the initial and
final states. Note that, for SDI if we use the experimental split of 138 keV instead of the
calculated split of 182 keV the lifetime would increase by a factor of 4 from 1.13 $\times$ $10^{-7}$
 s to 4.66 $\times$ $10^{-7}$ s. In either case the J=21/2$^+$ state is predicted to be isomeric.
 The results of the full space calculations are very close to the $g_{9/2}$ only case.

\subsection{Isomerism in $^{213}$Pb}

 In $^{213}$Pb, we have 5 neutrons in the $g_{9/2}$ shell. We are now at mid-shell. As explicitly
stated by Lawson \cite{2} we now have seniority as a good quantum number, 
secondly at mid-shell in the $g_{9/2}$ model space E2 transitions in which the seniorities do not change are forbidden.

\subsubsection{ \bf {Isomerism in the $g_{9/2}$ only space}}

Experimentally, the J=21/2$^+$ state, located at 1331 keV, will still have seniority v=3.
There are now 2 J=17/2$^+$ states below J=21/2$^+$ at 1141 keV and 1259 keV, respectively.
The experimental splits (energy differences) are 72 and 190 keV. In Table \ref{tab2} we can
clearly see the seniority selection rules coming into play. One of the B(E2)’s is zero
corresponding to a v=3 to v=3 transition. The other B(E2) is greater than the one in
$^{211}$Pb. The corresponding values are 60.28 and 25.19 $e^2 fm^4$. Here, we have a v=3 to v=5 transition corresponding to this large B(E2) value in $^{213}$Pb. The energy
difference in $^{213}$Pb is also larger than in $^{211}$Pb - 0.190 MeV vs 0.137 MeV. Both the
larger B(E2) and larger energy difference conspire to make the lifetime in $^{213}$Pb to be
shorter than that of $^{211}$Pb -  3.79 x 10$^{-8}$ s vs 4.66  x 10$^{-7}$ s.
 The problem is that this goes against experiment. The half life of $^{213}$Pb is larger than
that of $^{211}$Pb. The corresponding values are 0.26(2) $\mu s$ and 42(7) $ns$ .

\subsubsection{ {\bf Isomerism in the full space}} 

 There is a subtle change in gong from ``$g_{9/2}$ only" to full {\color{black} ($g_{9/2}$, $i_{11/2}$, $j_{15/2}$)}.
   First, we note there are 2 energy differences involved- a  large one (0.190  MeV )
and a small one (0.0712) MeV.
There are 2 basic transitions involved i.e. v=3 to v=3  forbidden; 
v=3 to v=5  strong.
In ``$g_{9/2}$ only" the v=3 to v=5 transition is associated with the large energy 
difference (0.190 MeV). The combination of large $B(E2)$ and large energy difference leads to a short lifetime.
In the full calculation, however  there is a switch. The large $B(E2)$ is associated with a small energy difference.   This leads to a longer lifetime than with
``$g_{9/2}$ only''. The respective values are 3.79 x $10^{-8}$ s for $g_{9/2}$ only and 6.48 x $10^{-6}$ s for full calculation,
but we must also consider for 
``Full" the small E2 transition (i.e.,  B(E2)=1.93 $e^2$fm$^4$). 

It is not zero in ``Full" because there is a small admixture (2.312 \%) of v=5 in a basic v=3 state. Given the 
large energy difference associated with this state ($17/2_1^+$) we get a lifetime
which is shorter than the one for the basic v=5 state ($17/2_2^+$) 1.18 x 10$^{-6}$ s.
This channel dominates the decay. 
%When all is said and done
The isomerism of the J=21/2$^+$ states is a complicated 
combination of seniority mixings and energy differences.
Using 1/T =1/T1+1/T2 we find that the half-life for the J=21/2$^+$ state in $^{213}$Pb is
9.982 x 10$^{-7}$ s.
In the Table \ref{tab_sen} we have shell model calculated seniority ratio for the $g_{9/2}$ and full model spaces. {\color{black} Here, seniority (v) is the number of nucleons not paired to angular momentum zero. Please see Ref. \cite{2} for more details.}

\subsection{Isomerism in $^{215}$Pb}

We might expect $^{215}$Pb (3 holes) to be very similar to $^{211}$Pb (3 particles). But there is a surprise
in the full case. The values of $B(E2)$ for $^{211}$Pb and $^{215}$Pb are respectively 25.35 and 10.59, respectively. {\color{black} The seniority v=3 is more dominant, although contributions from v= 3 and 5 are also coming. Although, experimentally only tentitive g.s. with ($9/2^+$) is known, we have reported shell-model results for few low-lyuing states.  The calculated first excited state $7/2^+$ is pushed up by 48 keV from  $^{213}$Pb to $^{215}$Pb in the full shell-model calculation.  The $21/2^+$ is lower in energy than $17/2_2^+$.}

\begin{table*}        
\begin{center}
%\begin{threeparttable}
\caption{ Shell model calculated seniority ratio for $^{211,213,215}$Pb isotopes corresponding to $g_{9/2}$ and full calculations. }
\label{tab_sen}
\begin{ruledtabular}
\begin{tabular}{ccc}
&$g_{9/2}$  & Full	\\[+1pt]
 \hline 
 $^{211}$Pb    & & \\
\hline
$21/2_1^+$ & {\bf v =3}  (100\%) &  {\bf v =3}  (100\%)  \\[+1pt]
$17/2_1^+$ &   {\bf v =3}  (100\%) &  {\bf v =3}  (100\%)  \\[+1pt] 
$17/2_2^+$ & - & {\bf v =3}  (100\%)  \\[+1pt] \\[+1pt]

 \hline 
 $^{213}$Pb    & & \\[+1pt]
\hline
$21/2_1^+$ & {\bf v =3}  (100\%) &  {\bf v =3}  (97.68\%) +  {\bf v =5}  (2.312\%)  \\[+1pt]
$17/2_1^+$ &  {\bf v =5}  (100\%) &   {\bf v =3}  (97.23\%) + {\bf v =5}  (2.765\%) \\[+1pt] 
$17/2_2^+$ & {\bf v =3}  (100\%) & {\bf  v =3}  (2.59\%) + {\bf v =5}  (99.74\%)  \\[+1pt] \\[+1pt]
\hline 
 $^{215}$Pb    & & \\[+1pt]
\hline
$21/2_1^+$ & {\bf v =3}  (100\%) &  {\bf v =3}  (78.44\%) + {\bf v =5}  (18.9\%) + {\bf v =7}  (2.6\%) \\[+1pt]
$17/2_1^+$ &  {\bf v =3}  (100\%) &   {\bf v =3}  (94.52\%) + {\bf v =5}  (2.79\%) + {\bf v =7}  (2.68\%)  \\[+1pt] 
$17/2_2^+$ & - &  {\bf v =3}  (77.20\%) + {\bf v =5}  (20.36\%) + {\bf v =7}  (2.4\%)   \\[+1pt] \\[+1pt]

\end{tabular}
\end{ruledtabular}
\vspace{-10pt}
\end{center}
%\end{threeparttable}
\end{table*}

\section {Summary}

In the present work we have addressed some properties of the low-lying levels in $^{211,213,215}$Pb isotopes using the single-$j$ ($g_{9/2}$) shell seniority concept. 
{\color{black} Shell-model calculation have been performed using KHH7B effective interaction alond with $QQ$ and delta interaction. The calculation corresponding to KHH7B interaction consist of two sets one with $g_{9/2}$ only, while full calculation with $g_{9/2}$, $i_{11/2}$, $j_{15/2}$ using $^{208}$Pb as a core.}

In Sec.II “the splitting of $E(21/2^+)$ – $E(3/2^+)$” we noted that
although the J = 21/2$^+$ and J = 3/2$^+$ states had definite seniorities, the splittings were in
general different for n=3 and n=5. They are the same only for seniority-conserving interactions.\\
 In Sec. III, we calculate the energies of 3/2$^+$ states in $^{211, 213, 215}$Pb. These states have not
been found experimentally. The calculations show the 3/2$^+$ states lying above those with
J=5/2$^+$ and 7/2$^+$. This partially explains why they have not been found experimentally.

In Sec. IV and V we discuss the isomerism of J=21/2$^+$ states in the three Pb isotopes. 
In ``$g_{9/2}$ only”, they all involve traditions between pure and unique
seniority states.
In $^{211}$Pb and $^{215}$Pb from J=21/2$^+$
v=3 to J=17/2$^+$ v=3 and in $^{213}$Pb from J=21/2$^+$ v=3 to 2 J=17/2$^+$ states - one with v=3 and the
other with v=5.

The experimental half life for $^{211}$Pb is smaller than that of $^{213}$Pb.  The respective values are 4.2 x 10$^{-8}$ s and  2.6 x 10$^{-7}$ s. For the full calculation, the respective numbers are 4.63 x 10$^{-7}$ s and 9.98 x 10$^{-7}$ s.

\section*{Acknowledgment}

We thank P.M. Walker for useful remarks.

\end{document}